\documentclass{aa}  

\usepackage{graphicx}
\usepackage{txfonts}
\usepackage{booktabs}
\usepackage{siunitx}
\usepackage{glossaries}

\newacronym{LBG}{LBG}{Lyman-break galaxies}
\newacronym{JWST}{JWST}{James Webb Space Telescope}
\newacronym{ALMA}{ALMA}{Atacama Large Millimeter/submillimeter Array}
\newacronym{HST}{HST}{Hubble Space Telescope}
\newacronym{SFRD}{SFRD}{star-formation rate density}
\newacronym{sed}{SED}{spectral energy distribution}
\newacronym{sfr}{SFR}{star formation rates}

\usepackage[pdfencoding=auto,psdextra]{hyperref}
\hypersetup{colorlinks=true, linkcolor=blue, filecolor=magenta, urlcolor=blue, citecolor=blue}
\usepackage[table]{xcolor}

\definecolor{apcolor}{HTML}{b3003b}

\definecolor{mkcolor}{HTML}{01abdf}

\begin{document}

\title{An optically-dark merging system at $z\sim6$ detected by JWST}


   \author{G. Rodighiero\thanks{email: giulia.rodighiero@unipd.it}
          \inst{1,2}
         \and
          A. Enia\inst{3,4}
          \and L. Bisigello\inst{5,1}
          \and G. Girardi\inst{1,2}
          \and G. Gandolfi\inst{6,7,8,2}
          \and M. Kohandel\inst{9}
          \and A. Pallottini\inst{9}\\
           N. Badinelli\inst{1}
         \and A. Grazian\inst{2}
          \and A. Ferrara \inst{9} 
          \and B. Vulcani\inst{2}
          \and A. Bianchetti\inst{1,2}
          \and A. Marasco\inst{2}
          \and F. Sinigaglia\inst{10}\\
       M. Castellano\inst{11}
       \and P. Santini\inst{11}
         \and P. Cassata\inst{1,2}
         \and E. M. Corsini\inst{1,2}
         \and C. Gruppioni\inst{12}
          }

   \institute{Department of Physics and Astronomy, Università degli Studi di Padova, Vicolo dell’Osservatorio 3, I-35122, Padova, Italy
        \and INAF - Osservatorio Astronomico di Padova, Vicolo dell’Osservatorio 5, I-35122, Padova
        \and University of Bologna - Department of Physics and Astronomy "Augusto Righi" (DIFA), Via Gobetti 93/2, I-40129, Bologna, Italy
        \and INAF - Osservatorio di Astrofisica e Scienza dello Spazio, Via Gobetti 93/3, I-40129, Bologna, Italy
        \and INAF - Istituto di Radioastronomia, via Gobetti 101, I-40129 Bologna, Italy
        \and
Scuola Internazionale Superiore Studi Avanzati (SISSA), Physics Area, Via Bonomea 265, I-34136 Trieste, Italy
\and
Institute for Fundamental Physics of the Universe (IFPU), Via Beirut 2, I-34014 Trieste, Italy
\and
Istituto Nazionale Fisica Nucleare (INFN), Sezione di Trieste, Via Valerio 2, I-34127 Trieste, Italy
\and
Scuola Normale Superiore, Piazza dei Cavalieri 7, I-56126 Pisa, Italy
\and
Département d’Astronomie, Université de Genève, Chemin Pegasi 51, 1290 Versoix, Switzerland
\and
INAF - Osservatorio Astronomico di Roma, Via Frascati 33, I-00078 Monteporzio Catone, Rome, Italy
\and 
INAF - Osservatorio di Astrofisica e Scienza dello Spazio (OAS), via Gobetti 101, I-40129 Bologna, Italy
\\
            }

   \date{Received September 15, 1996; accepted March 16, 1997}

 
  \abstract
   {Near- to mid-Infrared observations (from Spitzer and JWST) 
   have revealed a hidden population of galaxies at redshift $z=3$--$6$, called optically-dark objects, which are believed to be massive and dusty star-formers. They contribute substantially to the cosmic star-formation rate (SFR) density at $z\sim4$--$5$ (up to $30$--$40\%$).
   }
   {While optically-dark sources are widely recognized as a significant component of the stellar mass function, the history of their stellar mass assembly (and the evolution of their interstellar medium)  remains unexplored.
However, they are thought to be the progenitors of the more massive early-type galaxies found in present-day groups and clusters.   
   It is thus important to examine the possible connection between dark sources and merging events, in order to understand the environment in which they live.
   }
   {
   Here, we report our search for close companions in a sample of 19 optically-dark objects identified in the SMACS0723 JWST deep field. They were selected in the NIRCam F444W band and undetected below 2$\mu$m.
   We restrict our analysis to the reddest (i.e. F277W-F444W$> 1.3$) and brightest (F444W$< 26$ mag) objects.
    }
   {
   We have identified KLAMA, an optically-dark source showing a very close companion (angular distance $< \SI{0.5}{\arcsec}$). 
   The spatially resolved SED fitting procedure indicates that all components lying within \SI{1.5}{\arcsec} from the dark source are indeed at $z\sim5.7$. 
   Tidal features (leading to a whale shaped morphology) corroborate the hypothesis that KLAMA is the most massive ($\log (M_\star / M_\odot) > 10.3$) and dusty (A$_V\sim3$ at the core) system of an ongoing merger with a mass ratio of $\sim 10$. 
   Merging systems with properties similar to KLAMA are identified in the \texttt{SERRA} simulations, allowing us to
   reconstruct their stellar mass assembly history and predict their molecular gas properties 
   } 
   {The discovery of mergers within dark galaxies at the end of the Epoch of Reionization underscores the importance of conducting a statistical search for additional candidates in deep NIRCam fields. 
   Such research will aid in understanding the significance of merging processes during the obscured phase of stellar mass accumulation. 
}
   \keywords{--
                --
               }

   \maketitle
%

\section{Introduction}
\label{intro}


While the formation of stars in cosmic structures has been a pivotal observational focus for many decades, modern astronomical facilities are continuously revising the timeline for the emergence of the first galaxies. 
 In this quest, \gls{HST} has been a pioneering force, particularly in its exploration of high-redshift galaxies through their rest-frame ultraviolet (UV) emission. These high-redshift galaxies, called \gls{LBG} for their distinctive selection method, have undergone extensive scrutiny, spanning from approximately $z \sim 3$ to $z \sim 11$ \citep[e.g.][]{Reddy09,Mclure13,Finkelstein2015,Oesch2016,Ono2018,Atek2018}. 
  These galaxies typically exhibit moderate \gls{sfr} and stellar masses, making up a significant portion of the galaxy population. 
 
However, despite thorough research on these ‘normal’, unobscured galaxies, the full understanding of the entire galaxy population remains incomplete beyond $z > 3$. Such incompleteness arises because systematic selections based on rest-frame UV often miss massive, obscured sources. In addition, the physical properties of these obscured sources, such as redshift, stellar mass and \gls{sfr}, are often difficult to constrain from the available \gls{sed}.

Such bias also affects the measurement of the \gls{SFRD} of the Universe \citep[e.g.,][]{madaudickinson}. The \gls{SFRD} reaches its peak around $z \approx 1$-–$3$, an era commonly referred to as the ‘Cosmic Noon’, and then experiences a rapid decline up to the present epoch. However, numerous investigations indicate that the fraction of the \gls{SFRD} obscured by dust, not considered in optical/UV surveys at $z > 2$, is likely significant (up to $30$--$40\%$) and increases with redshift, at least up to $z \approx 5$-–$7$ \citep[e.g.,][]{Novak_2017,Gruppioni_2020,Jones2020,Fudamoto21,Barrufet,Fujimoto23,Algera23,Xiao23b,Barrufet_LF}. 

A significant population of optically-dark galaxies (ODGs) emitting relatively bright infrared (IR) or sub-millimeter (sub-mm) radiation has been discovered in {\it Spitzer} \citep{Werner} data. Some ODGs have been detected using the Atacama Large Millimeter/submillimeter Array \citep[ALMA, e.g.][]{Wang2016,Xiao23a,Xiao23b} and radio telescopes \citep{Talia2021,Enia2022}. These galaxies have very red colors and remain undetected even in deepest \gls{HST}-band observations
\citep{Rodighiero2007,Sun2024}.

The {\it James Webb Space Telescope} is now transforming our understanding of optically dark objects, enabling the detection of galaxies much fainter than $Spitzer$-detected ones at $z > 3$ (i.e. [F444W]$> 25 $ mag). Various NIRCam colour combinations have been proposed to identify such massive and dusty systems, in particular through an F277W-F444W cut. The reddest selection (e.g. F277W-F444W$> 1.5$) has proved to include a “triality” \citep[as in][]{PPG23} of different populations: 1) red AGNs at $4 < z < 8$ (called “little red dots” for their point like morphology) that, however, show are flat in the UV and steep in the optical rest frame of their \gls{sed} \citep{labbe_LRD,Furtak_LRD,Greene23,Kokorev24,PPG24,Kocevski}; 2) passive sources dominated by old stellar populations at $3 < z < 5$ \citep[e.g.][]{Carnall23,Valentino23}; 3) massive star-forming systems at $3 < z < 8$  \citep{Akins23,Barrufet,Rodighiero23,Gottumukkala,Barro23}. The latter has been confirmed to provide a dominant contribution to both the obscured \gls{sfr} density at $3<z<8$ \citep{Williams23} and to the high-mass end of the stellar mass function at similar epochs \citep{Rodighiero2007,Gottumukkala}. 

The clustering properties of ODGs at $z \sim 4$ \citep[in particular for a sample of dusty $H$-dropouts,][]{wang19} suggests that some of these objects could reside in the most massive dark matter halos at the epoch of observation, making them likely progenitors of the biggest early-type galaxies in today's groups and clusters. 
The current consensus on the formation and evolution of cluster cores and Brightest Cluster galaxies (BCG) is that their stellar mass forms early ($z > 4$) in separate galaxies \citep{DeLucia2006} that then eventually assemble the main structure at late times ($z < 1$). 
Understanding how these massive galaxies form and evolve over time and what regulates and eventually shuts off their star formation are open questions in the field of galaxy evolution.

At Cosmic Noon and beyond, some studies \citep[e.g.][]{Clements14,Dannerbauer14,Casey2016,Kato16} report an increase in the \gls{SFRD} on the order of magnitudes, when compared to the average field values at the same $z$, in 
 the volumes filled by the cores of 
 dusty protoclusters.
  Moreover, a blind [CII] line search at $z \sim 5$ \citep[ALPINE survey,][]{Loiacono21} found that, for sources residing in overdense regions,  the resulting   \gls{SFRD} is ten times higher than the corresponding the field value. Similar results were also obtained for sources serendipitously detected in
the continuum by ALMA \citep{Gruppioni_2020}. These hints of enhanced \gls{SFRD} in highly dense regions could 
be driven by larger gas/stellar mass content of clustered dusty galaxies
and by environmental processes \citep{Lemaux}.

A systematic study of the clustering properties of optically-dark galaxies in the \gls{JWST} era is still missing. However, the role of the environment can be currently investigated by studying the presence of local overdensities around such massive systems. In particular, the role of galaxy mergers and interactions in the growth and evolution of the reddest galaxies is fundamental to our understanding of the Universe.
Interactions among galaxies are crucial in shaping their physical characteristics and significantly contribute to their mass accumulation and growth, leading to morphological changes and frequently culminating in mergers \citep{Toomre72}. 
At high redshifts, major mergers are expected and observed to increase their significance, with their fraction scaling as  $(1 + z)^{2-3}$ up to $z \sim 5$--$6$ \citep[e.g.][]{Rodriguez-Gomez2015,Ventou2017,Snyder2017,Romano21} and approaching a constant value at higher redshifts \citep[e.g.][]{Dalmasso}.


Many merger systems have already been identified at $z>5$--$6$ with \gls{HST}, \gls{JWST} and ALMA. These include some Ultra Luminous IR galaxies at $z \sim 7$ \citep[e.g.][]{Hygate2023} and UV-bright Lyman $\alpha$ emitters at $z \sim 6.6$ \citep[e.g.][]{Ouchi2013,Boyett24}. However, direct detection of a merger with high-resolution optical imaging associated with a classical optically dark galaxy has not been reported yet \citep[but see the "Jekyll \& Hide" analogue case at lower redshift,][]{Schreiber18}. This work is thus a pilot study aimed at characterizing the physical condition of the candidate group. A systematic search of mergers, including even fainter ODGs, is necessary to probe their statistical impact in the assembly of massive sources. 

Here, we report the discovery of a candidate multiple merger at $z \sim 6$, involving one of the F444W/NIRCam brightest dark objects identified in the early \gls{JWST} observations in the SMACS0723 field. The presence of a close blue companion, combined to the evidence of some tidal features, classify this source \citep[called KLAMA, see][hereafter R23]{Rodighiero23} as a group of galaxies whose stellar mass content is dominated by the optically-dark system. While this system is quite unique among the public \gls{JWST} extragalactic fields, its existence highlight the importance of environmentally-driven processes in feeding the stellar mass assembly (such as galaxy interactions) of the most obscured systems when the Universe was just one billion years old \citep[see also][]{gentile24}.
This works is thus a pilot study, aimed at characterizing the physical properties of the candidate group. A systematic search for mergers, including even fainter HST-dark objects, is necessary to probe their statistical impact in the assembly of massive sources. 

The paper is structured as follows. In Sect. 2 we describe the \gls{JWST} dataset adopted for our analyses. In Sect. 3 we report the selection of optically-dark systems presenting a close companion, while Sect. 4 describes the spatially resolved study that we adopt to derive the physical properties of the sample. The main results are reported in Sect. 5, and compared to hydrodynamical simulations in Sect. 6. Thorough the paper, we assume a $\Lambda$CDM cosmology with parameters from \cite{PlanckCollaboration2016} and a \cite{Chabrier_2003} Initial Mass Function (IMF).\

\section{NIRCam \gls{JWST} observations of SMACS0723}


The initial images of SMACS0723, obtained using the NIRCam and MIRI instruments, showcase the exceptional quality of \gls{JWST}'s observations \citep{2022arXiv220713067P}. The Early Release Observations were designed to confirm \gls{JWST}'s capability to capture high-redshift galaxies with unparalleled depth, surpassing the capabilities of \gls{HST} and ground-based telescopes. SMACS0723 is one of the 41 most massive galaxy clusters identified by {\it Planck} at redshifts around $z\sim 0.2$--$1.0$, a target of the \gls{HST} program RELICS (Reionization Lensing Cluster Survey).


The NIRCam instrument directed its focus on SMACS0723, allocating one detector to the cluster and the other to an adjacent off-field region. Exposures were conducted with NIRCam filters F090W, F150W, F200W, F277W, F356W, and F444W, each lasting approximately 7500 seconds. This resulted in a $5 \sigma$ sensitivity limit of approximately $28.5$--$29.6$ AB magnitude for point-like sources. These depths are comparable to those achieved with WFC3/IR for the HUDF12 pointing \citep{Koekemoer}, and they exhibit a sensitivity level ten times greater than the deepest {\it Spitzer}/IRAC imaging available at 3.6 and 4.5 $\mu$m. NIRCam reduced images are available from the Mikulski Archive for Space Telescopes (MAST).\footnote{\url{https://archive.stsci.edu}.} In this work, we adopt the NIRCam mosaic produced with the GRIZLI pipeline \citep{Brammer2021}.\footnote{\url{https://erda.ku.dk/archives/7166d013c1ca1371aac3c57b9e73190d/published-archive.html}}

\begin{figure}
    \centering
    \includegraphics[width=7cm]{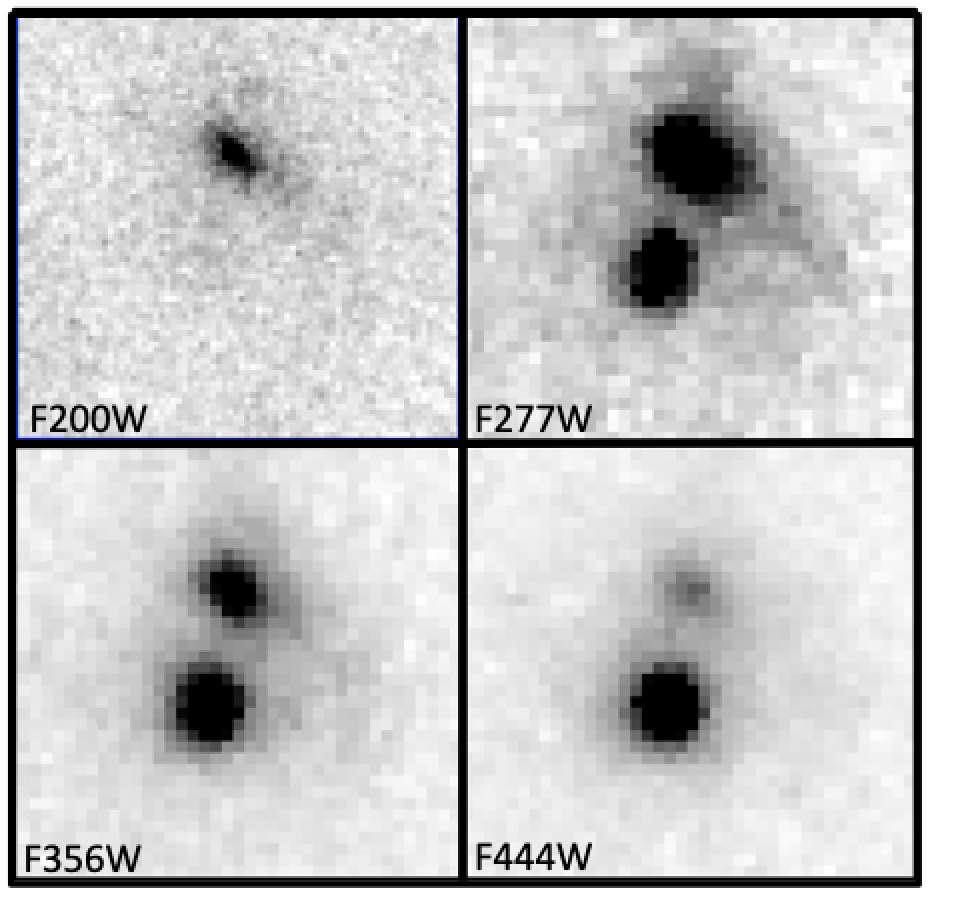}
    \caption{NIRCam cutouts for KLAMA, including filters F200W, F277W, F356W and F444W (from top-left to bottom-right, respectively; North is up, East at left). The size of each postage stamp is $\SI{1.5}{\arcsec} \times \SI{1.5}{\arcsec}$. KLAMA is the southern source, clearly undetected in the F200W band. A bluer companion is located at \SI{0.4}{\arcsec} in the northern direction. The nature of the two objects is further discussed in the paper.}
    \label{fig:klama-sel}
\end{figure}

\section{Searching for mergers among optically dark objects in the SMACS0723 field}\label{selection}

\subsection{Mergers selection}
\begin{figure*}
    \centering
     \includegraphics[width=12cm]{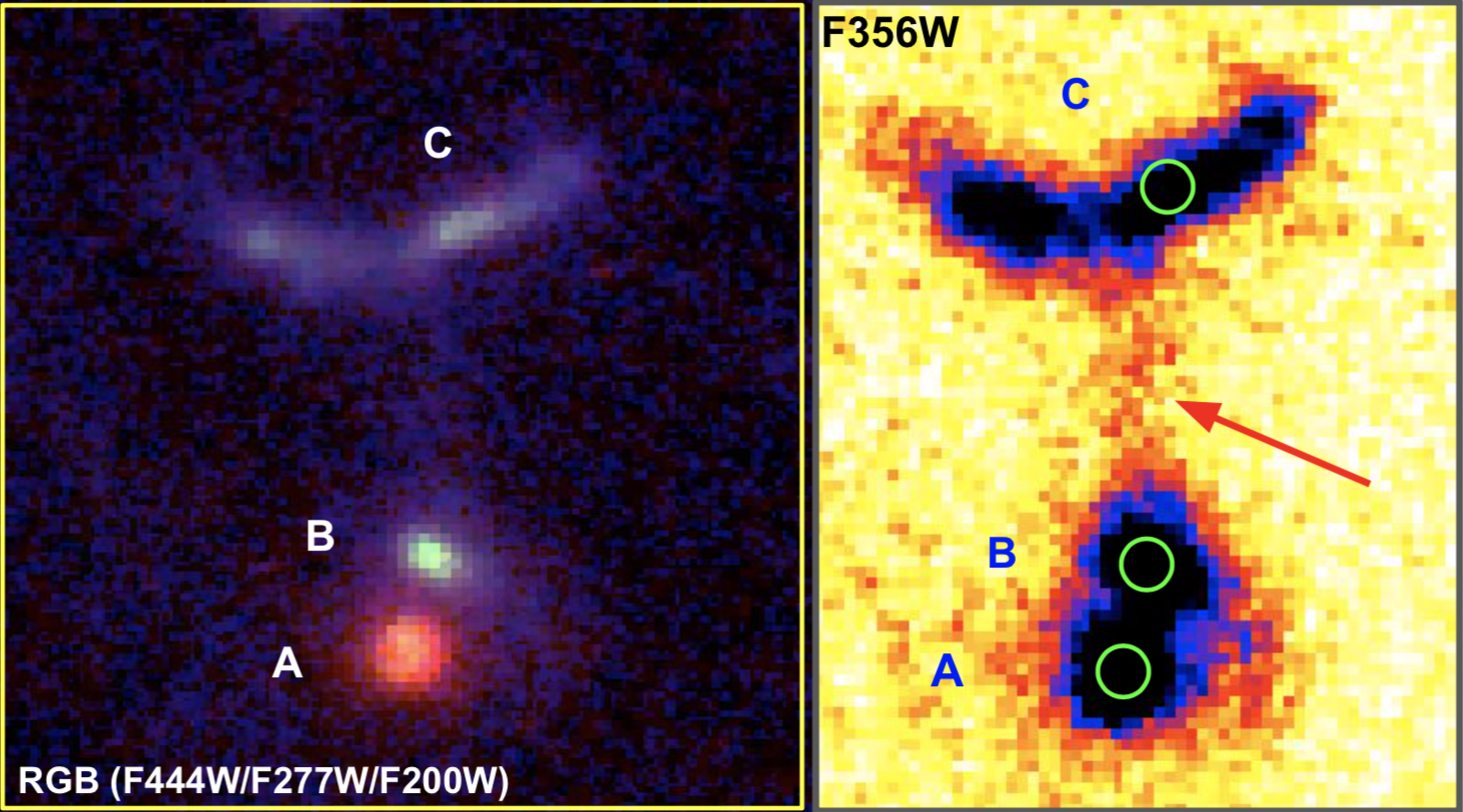}
    \caption{{\it Left panel}: RGB image of the KLAMA system (Red=F444W, Green=F277W, Blue=F200W). For reference, the yellow square marks 
    {\bf  a ($\SI{3}{\arcsec} \times \SI{3}{\arcsec}$) area}. We indicate the three components of the galaxy group with capital letters (A, B, and C). KLAMA is the reddest source (A).
    {\it Right panel}: F356W highly contrasted image of the same region shown in the left panel. The three components are indicated. Interestingly, a diffuse emission clearly rises as a bridge (marked with the red arrow) connecting the main system (A+B) to the tail (C).
    The green circles represent the apertures (equal to the FWHM of the F444W PSF) and the positions used to extract the photometry for the \texttt{Bagpipes} fit.  
    }
    \label{fig:klama-rgb}
\end{figure*}

We explore the existence of close companions within a set of optically dark galaxies, selected to be candidate massive and dusty sources at $z > 3$. 
We focus on a pilot sample of 19 sources, presented in \cite{Rodighiero23}, that we derived from the Early Release Observations (ERO) of the SMACS0723 field, \citep{Pontoppidan}.
The selection was based on a catalogue of F444W detections at the 5$\sigma$ limit of [F444W] $< 29$ mag, lacking an F200W counterpart in blind photometric catalogues obtained with \texttt{SExtractor}. Such dropout technique
naturally classifies these sources as optically dark objects. 
We note that, after selection, we performed ad hoc photometry
on each object of the sample. This refined procedure, better accounting for the local variation of the sky background, has turned 
some non-detections at wavelengths shorter than F200W in the preliminary \texttt{SExtractor} catalogs into faint detections (see R23 for details). 


As in this study we are mainly interested 
in identifying merging processes in 
massive dust obscured sources at $z>3$, 
we limit our analysis to the reddest objects. To this aim, we apply a conservative color cut of F277W-F444W$> 1.3$ mag in order to maximize the probability of identifying high-$z$ dusty sources. We relax more extreme criteria like those adopted by \citet[][F277W-F444W $> 1.8$ mag]{Akins23} or \citet[][F277W-F444W $> 1.5$ mag]{Barro23} in order to account for the scatter due to the photometric uncertainties in the NIRCam bands.
We also impose a flux cut of [F444W]$ < 26$ mag (roughly corresponding to a stellar mass of $\log (M_\star / M_\odot) = 10.5$ at $z \sim 6$) to limit our investigation to the most massive candidates.
Among the 19 sources of R23, seven objects have color redder than F277W-F444W $> 1.3$ (with values spanning the range $1.3$--$5$ mag), but only one source stands with a magnitude [F444W] $< 26.00$ (KLAMA, as dubbed in R23). 

Here, we neglect the effect of lensing magnification, since the targeted source is located 
at \SI{1.4}{\arcmin} of distance from the cluster core. In support of this assumption, we note that \cite{Castellano16} has shown that the parallel observations of {\it Hubble} Frontiers Fields have little magnification (e.g. 10-15\% at most).

The NIRCam images of the selected source are presented in Figure \ref{fig:klama-sel}. In the next sections we provide an in depth investigation of the morphological and spatially resolved photometric properties of KLAMA and of its local environment.


\begin{figure*}
    \centering
    \includegraphics[width=14cm]{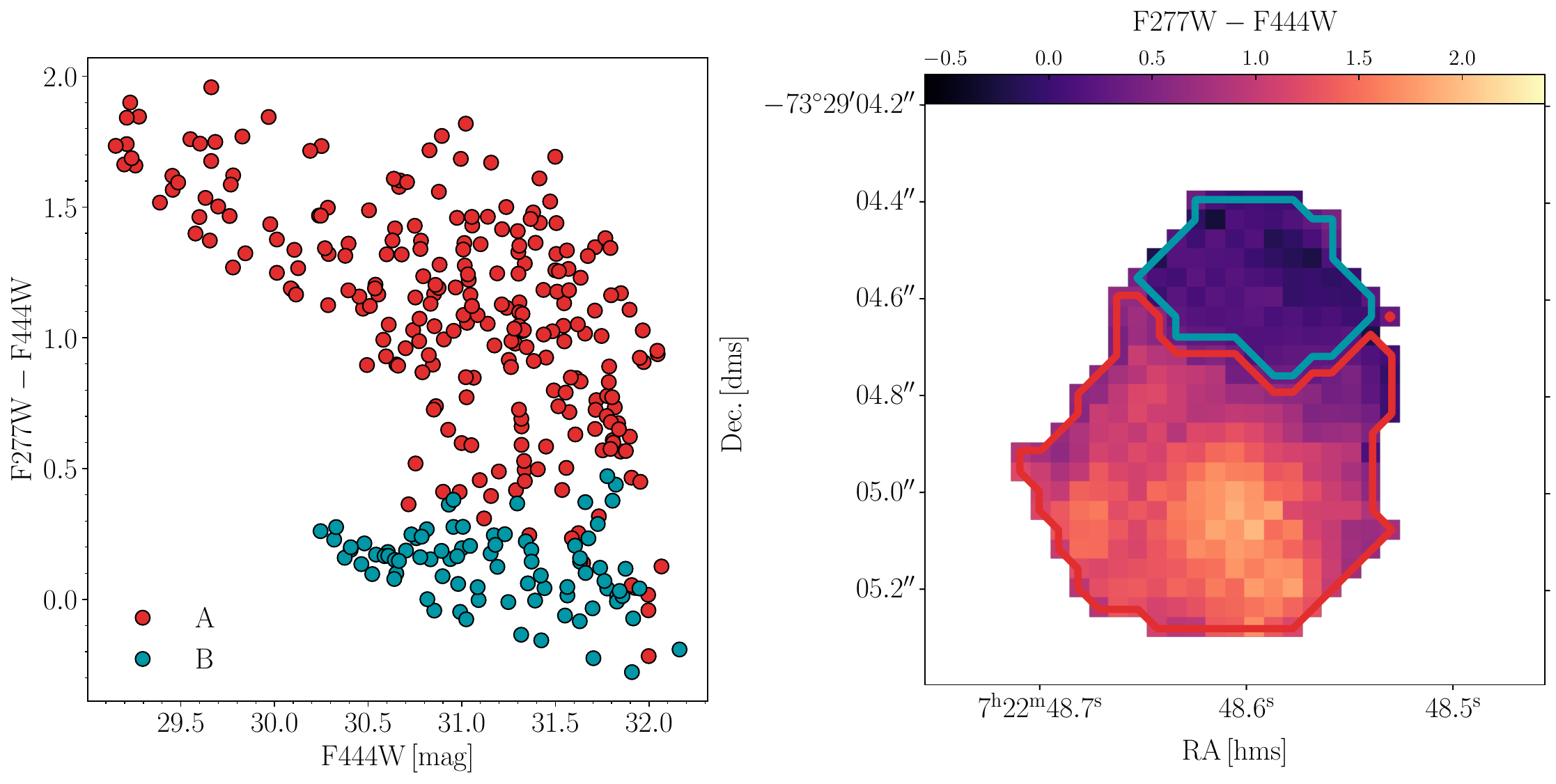}
    \caption{{\it Left panel:} Spatially resolved color magnitude diagram for KLAMA and its closest companion (components A/red  and B/blue, respectively). Each filled circle corresponds to a pixel of the parent source, as detailed in the right panel. {\it Right panel:} F277W-F444W color map for sources A and B. The color scale highlights the gradient from KLAMA towards the companion. The spatial separation between the two sources is based on a \texttt{SExtractor} segmentation map. The boundaries are marked with red and blue solid lines for components A and B (respectively). The pixel size is that of the original NIRCam detector.}
    \label{fig:klama-colcol}
\end{figure*}

\subsection{KLAMA: a candidate merging system}
As clearly visible in Fig.\,\ref{fig:klama-sel}, KLAMA is an F200W dropout source, resolved in all bands where it is detected.  
A nearby source is found around $\SI{0.4}{\arcsec}$ from the galaxy center, towards the northern direction. 
Since we are interested in identifying potential mergers among optically dark sources, it is very relevant to understand if the two objects in the small field of view ($\SI{1.5}{\arcsec} \times \SI{1.5}{\arcsec}$) of Fig.,\ref{fig:klama-sel} are physically associated or if they simply sit along the same line of sight.

Thus, it is interesting to zoom out from the small region shown in Figure \ref{fig:klama-sel}, to better characterize the environment in a larger volume around the potential double system.
This is presented in Figure \ref{fig:klama-rgb} (left panel), where we
report a NIRCam color map (RGB image combining F200W, F277W and F444W filters)
in the immediate surrounding region, i.e. $\SI{3}{\arcsec} \times \SI{3}{\arcsec}$, to look for the presence of other potential companions.

KLAMA stands as the reddest object in the bottom part of the field (A component) with its bluer close companion (B component). This snapshot reveals the presence of an elongated structure in the upper part of the map (which we mark as the C component). While the blue colors of B and C might naturally indicate a different redshift as compared to the red A source (KLAMA), in the next section, we will provide evidence for a physical connection among the three components.
Once the photometric analysis of this system will be presented in Section \ref{sec:photoz_results}, we will further discuss the eventual presence of large overdensities in the SMACS0723 field at the same photo-$z$ of the KLAMA group.

To enhance the color bimodality among objects A and B, we report in Figure \ref{fig:klama-colcol} (left panel) a spatially resolved color-magnitude diagram (i.e. F277W $-$ F444W versus F444W) of the pair. Red and blue-filled circles correspond to the values of each pixel (at the original size of the NIRCam detector pixels) of each source (A and B, respectively). 
The spatial separation between the two sources is based on
a \texttt{SExtractor} segmentation map.
The right panel shows the distribution of the F277W-F444W color in the sky plane. Clearly, the central part of KLAMA turns to be up to two magnitudes redder than its potential companion, reaching values as red as those of the very dusty galaxies at $z > 4$ \citep[e.g.][]{Akins23, Bisigello23,Barro23,Williams23}.

\begin{figure}
    \centering
    \includegraphics[width=9cm]{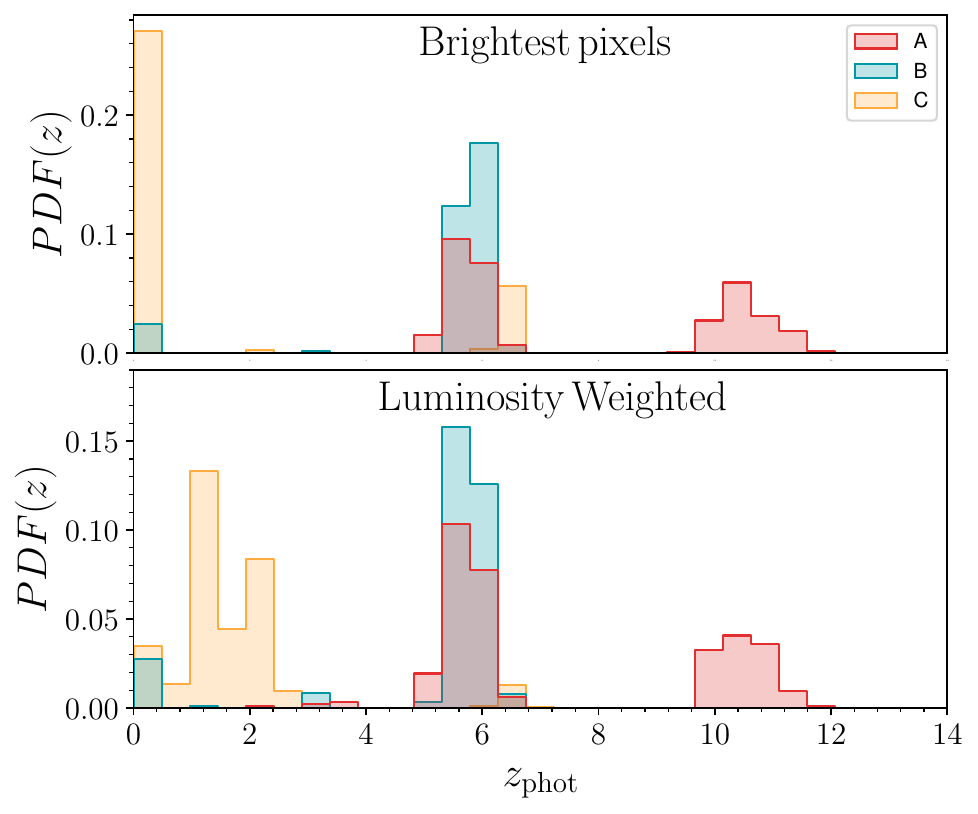}
    \caption{{\it Top panel - brightest pixels:} Probability Distribution Functions of the photometric redshift solutions derived by fitting with \texttt{Bagpipes} the brightest convoluted pixel of each source (A, B and C), at the positions indicated by the green circles in the right panel of Fig. \ref{fig:klama-rgb}. Each of the three components (A, B and C)  is represented with a different color, as marked in the legend.
    {\it Bottom panel - all pixels:} Luminosity weighted PDF of the redshifts from all the (convoluted) pixels belonging to a source.}
    \label{fig:pdf}
\end{figure}


\section{Methodology}\label{Sedfitting}
The superb angular resolution achievable with the \gls{JWST} allows us to probe the optical morphology of galaxies up to $z \sim 6$--$8$, translating into physical scales of $0.7$--$0.5\ {\rm kpc}$. In this work, we maximize the photometric information, appreciable in Fig.\,\ref{fig:klama-sel}, by adopting a spatially resolved \gls{sed}-fitting approach to study the KLAMA system 
(for clarity, in the following, we will refer to KLAMA as only to the A component, while we will call {\it the Whale} the whole system composed by A+B+C. 
This method has already been successfully applied to estimate photometric redshifts and stellar population properties on a pixel-by-pixel basis in \gls{JWST} samples (see \citealt{PPG23} and \citealt{Argumanez} for a detailed methodological discussion; see also \citealt{Abdurrouf}).

\subsection{Source segmentation}
From a visual inspection of the RGB image (left panel of Fig.\,\ref{fig:klama-rgb}), it is clear how the system can be decomposed into three components:
\begin{itemize}
    \item A, the southernmost redder component (i.e. KLAMA);
    \item B, its companion, with bluer colors; 
    \item C, a more extended uppermost component, which could in principle be subdivided into other two components reminiscent of two highly inclined spiral galaxies.
\end{itemize}
We recover the same results of the visual inspection by applying the Source Extraction and Photometry library \citep[\texttt{SEP},][]{Barbary2016}, a Python-wrapper of \texttt{SEXtractor} \citep{1996A&AS..117..393B}, to the available maps. This segmentation is then applied in the following Sections in describing individual properties of the sources composing the system.

\subsection{Spatially resolved SED fitting}
In the following paragraph, we briefly summarize our multiwavelength spatially-resolved procedure, while we demand a more accurate description in a forthcoming publication (Enia et al., in prep.). The full procedure follows three steps:
\begin{itemize}
\item[(i)] degradation of each image to a spatial resolution equal to the lowest available (in our case, that provided by the F444W band, i.e., 0.14"); 
\item[(ii)] measuring the flux and associated error in every pixel for each filter;
\item[(iii)] \gls{sed}-fitting, deriving the physical properties on the scale of individual pixels.
\end{itemize}

\subsubsection*{PSF-matching and photometry}
The first step is to degrade the resolution of all the available maps to the lowest one. In this case we have six filters, ranging from $0.9\,\mu$m to $4.4\,\mu$m, with the lowest one being the F444W filter. We thus convolve all the other filters using the convolution kernels built following \cite{2011PASP..123.1218A}\footnote{Available here: \url{https://www.astro.princeton.edu/~draine/Kernels.html}}, matching each observation with the PSF of F444W. When dealing with weight images (or error ones) we perform the convolution step with the square of the convolution kernels.

All the images are at the same pixel scale (\SI{0.04}{\arcsec}), so there is no need to re-bin them to match the scale. However, we group together pixels into $3 \times 3$ pools to match the PSF of the F444W image, summing (or square-summing, for error images) the single pixel fluxes within the pool. This is to reduce the impact of information correlation between close pixels due to the PSF-convolution smearing. In the end, the sampled scales are of \SI{0.12}{\arcsec} which correspond to physical scales of $\sim 0.7\,\mathrm{kpc}$ at $z \sim 6$.

At the end of this step, we have a three dimensional datacube, containing in the spaxels all the fluxes at each ($3 \times 3$ pool of) pixels per each band.

\begin{figure*}
    \centering
    \includegraphics[width=18.5cm]{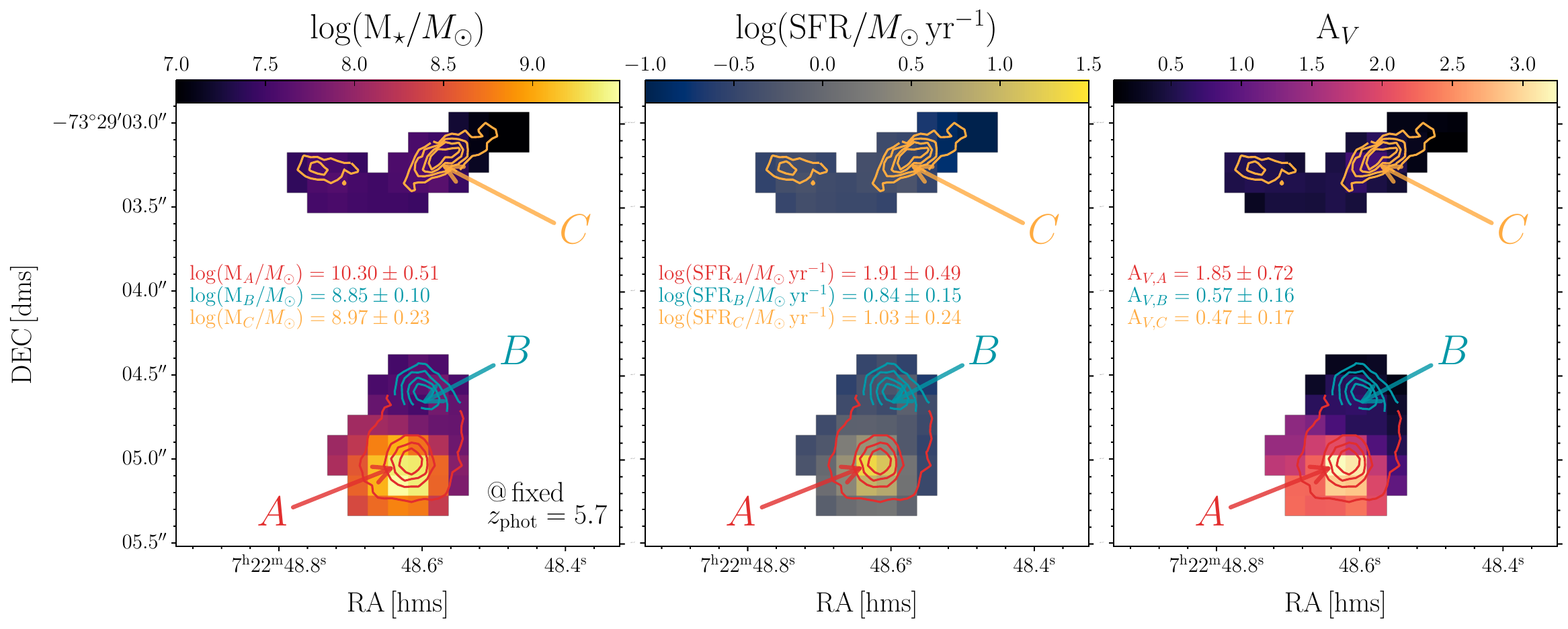}
    \caption{Spatially resolved outputs of the \texttt{Bagpipes} run obtained by fixing the redshift of all pixels in the system at $z=5.7$. The segmentation map obtained from \texttt{SEP} is used to identify the physical regions associated to each of the three main components (A, B and C) identified in Fig. \ref{fig:klama-rgb}. In the three plots, from left to right, we report the spatial distribution of stellar mass, \gls{sfr} and $A_V$, respectively. The pixel size used in this map accounts for the convolution with the PSF in each NIRCam band. In this figure we adopt the F444W PSF as reference.  In each panel we report the integrated values of the physical properties obtained for the three separate components. Contours of the F444W flux emission are reported as solid lines in each plot.}
    \label{fig:physical}
\end{figure*}

\subsubsection*{SED-fitting with \texttt{Bagpipes}}
We perform \gls{sed}-fitting with \texttt{Bagpipes} \citep{2018MNRAS.480.4379C}, an algorithm widely used in extragalactic astronomy \citep{2021ApJ...910..135S}, in recovering the photometric redshifts and physical properties of the first detected sources in the premiere \gls{JWST} public releases \citep{2023ApJ...956...61A, 2023MNRAS.520.4554D, 2023MNRAS.520.3974C, 2023Natur.619..716C}, and also for spatially-resolved analysis \citep{2023ApJ...948..126G}. At its full capabilities, \texttt{Bagpipes} is able to simultaneously deal with spectral and broad band photometric information coming from the same source. Here, we only deal with the latter,  as no spectroscopic information are available at the moment. We only fit the spaxels belonging to the above-mentioned A, B, and C components obtained after image segmentation, thus ensuring that a satisfying level of signal-to-noise ratio is achieved (i.e. $S/N>3$).

Differently from the released version of \texttt{Bagpipes}, we refine the definition of the likelihood in the code in order to properly account for upper limits to our photometric measurements, that are present in the outskirts of KLAMA in the shorter-wavelength filters, following the recipe in Eq.\,8 of \cite{2012PASP..124.1208S}. 

Following  \cite{Rodighiero23}, we fit the star formation histories with an exponentially declining $\tau$-model, $\mathrm{SFH} \propto \exp^{-t/\tau}$, with uniform priors on $\tau$ between $0.01$ and $10$ Gyr and time since the beginning of the star formation between $0.001$ and 14 Gyr. All of these -- as the following ones -- are uniform priors. Stellar metallicity is left free to vary between 0.005 and 2.5 solar metallicities, and the total mass formed priors are $1.0 < \log (M_{\star} / M_\odot) < 12.5$. We also include a nebular component with logarithm of the ionization parameter varying between $-4$ and $-2$. As for the dust, we use a Calzetti attenuation law \citep{2000ApJ...533..682C} with $A_V$ priors between $0$ and $6$. The spaxels photo-$z$ is left as a free parameter, varying between $0$ and a maximum of $15$, except in the run where we fix the redshift of all the components to $5.7$ (see Sect.\,\ref{sec:photoz_results} for further details).

From the \texttt{Bagpipes} provided PDFs, we take the $16$th, $50$th and $84$th percentiles to estimate the fiducial value and errors of the parameters. We also verified that the results are consistent when taking the maximum-likelihood estimator from the PDFs. In the end, we have a catalog of all the recovered photo-$z$s and physical properties (as stellar masses, SFR, dust attenuation) of each spaxel, from which we can start our data analysis.

\section{Results}

\subsection{Photometric redshift of the system}\label{sec:photoz_results}
We have run \texttt{Bagpipes} over all beam convoluted pixels, in a region covering the whole system. The most relevant information is the determination of the photometric redshift for the three components, in order to statistically evaluate the likelihood that they are, indeed, a physically bound system.

We firstly consider the results drawn from lux measurements extracted in a fixed aperture centered on the brightest pixel in each of the three components. The
angular size (corresponding to the FWHM in the F444W filter) and the positions of these three apertures are reported as green circles in Fig. \ref{fig:klama-rgb} (right panel). For components A and B they coincide with the luminous centers of
their host galaxies, while for the elongated structure C it corresponds to the  position of a likely star-forming clump.
These apertures represent the less contaminated regions of each component (in particular for  A and B), in terms of flux from nearby objects polluting the surrounding environment.

In Fig.\,\ref{fig:pdf} (top panel) we report the PDF($z$) for these three small regions. KLAMA (red histogram) shows a bimodal distribution, with a primary solution around $z \sim 5.7$ and a secondary peak at $z \sim 10$ (associated to the position of the Balmer- or Lyman-breaks, respectively). As we will see, the identification of such a bright source at $z > 10$ would correspond to a very massive galaxy ($M_\star > 10^{11} M_\odot$), challenging standard cosmological models \citep[e.g.][]{Menci22,Boylan,Lovell23}.
The B component (blue histogram) has a very narrow and peaked solution at the same low-$z$ solution of KLAMA, i.e. $z \sim 6$.
This preliminary result suggests that A and B could formally be a physical pair of sources sitting in the same cosmic volume 
(which would corresponds to a physical distance between their centers of $\sim2.5$kpc). 

\begin{figure}
    \centering
    \includegraphics[width=9cm]{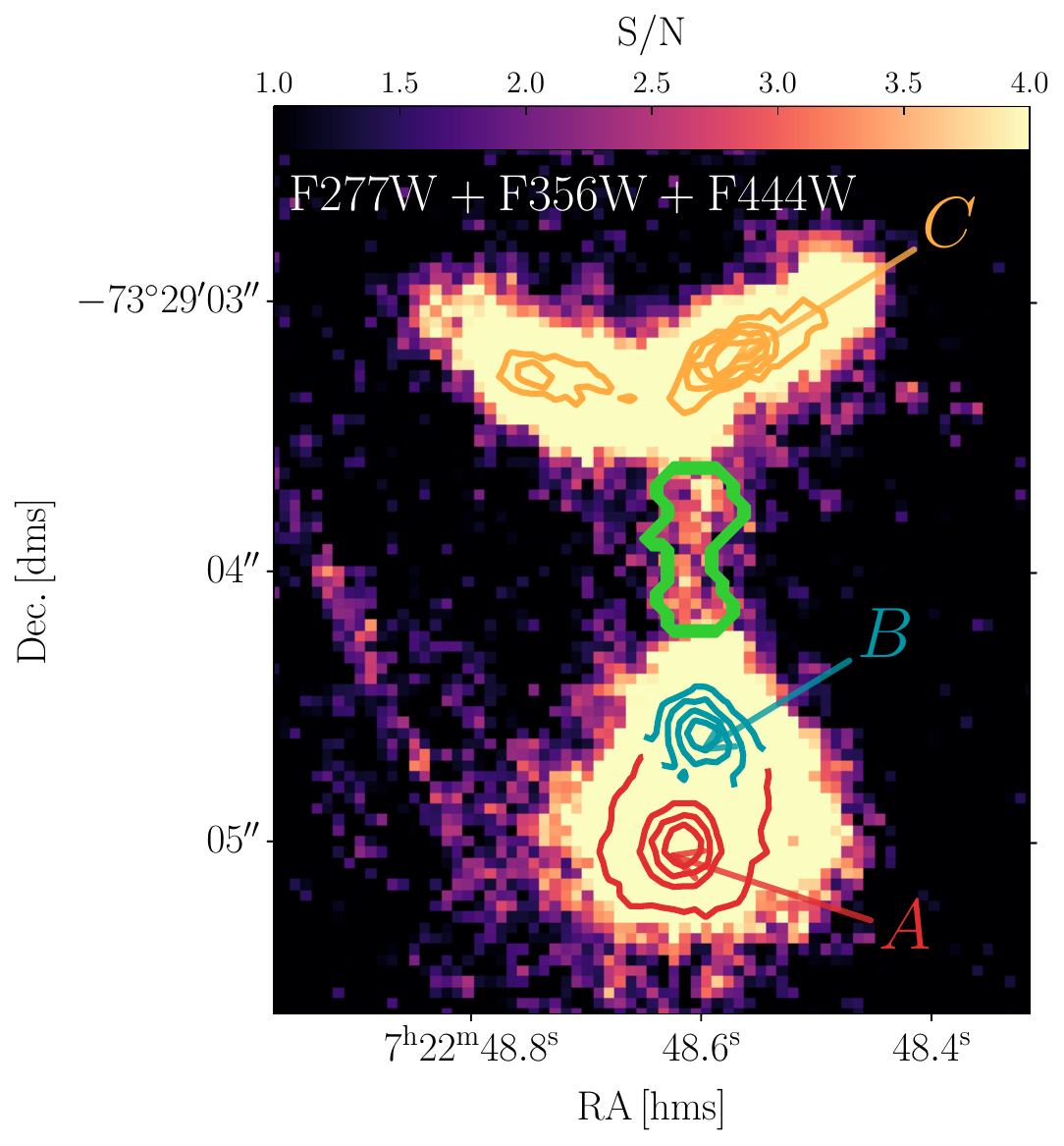}
    \caption{The figure shows the S/N spatial distribution of the candidate merging system, derived by coadding three NIRCam maps (F277W, F356W, F444W), after convolution to the longest wavelength. The colour bar marks the range of S/N variation across the pixels. The solid lines indicate contours of equal S/N for the three components of the system (red: A, light blue: B, orange: C). Five equally spaced contours are shown, starting from a S/N of five up to the maximum. The diffuse, tidal emission between B and C presents various pixels with S/N $>3$, and an average S/N $\sim2.85$ (computed within the region delimited by the green solid line). }
    \label{fig:bridge}
\end{figure}

\begin{figure*}
\includegraphics[width=0.32\textwidth]{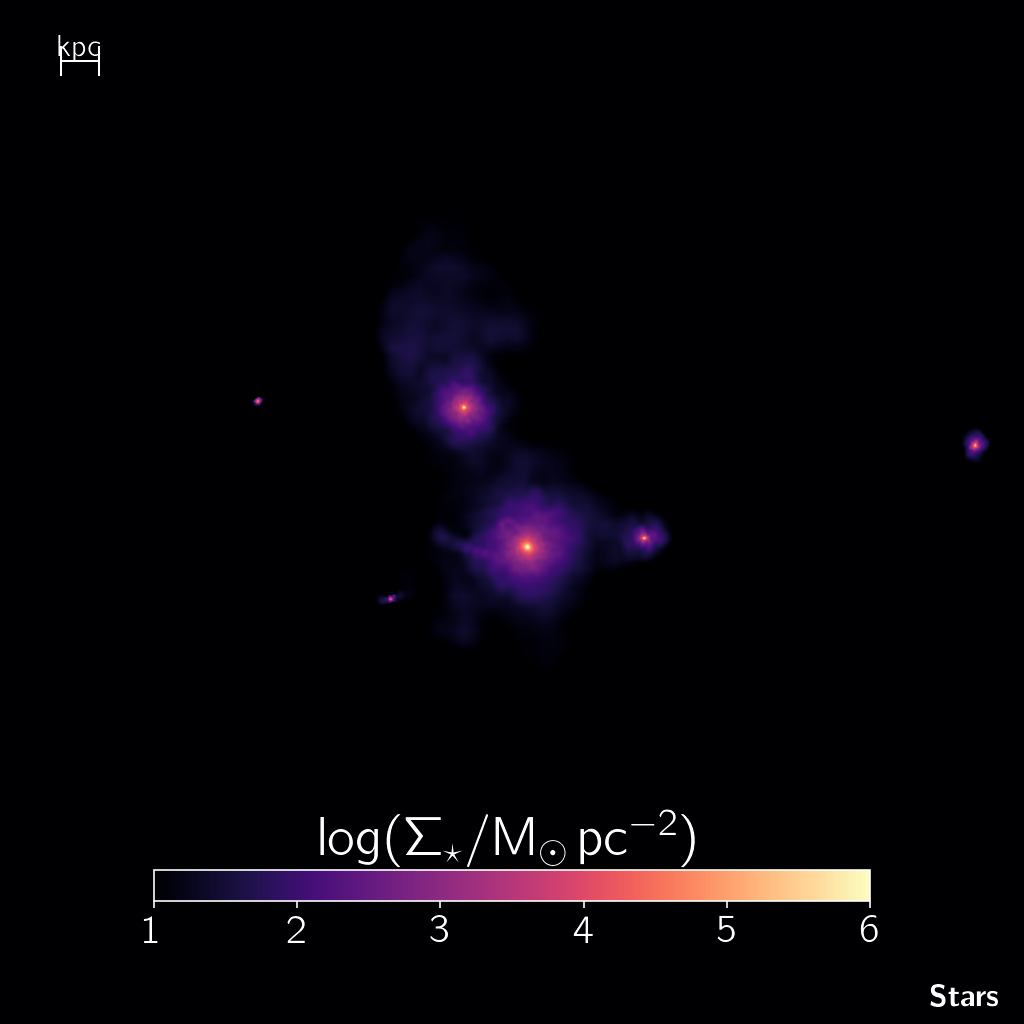}
\includegraphics[width=0.32\textwidth]{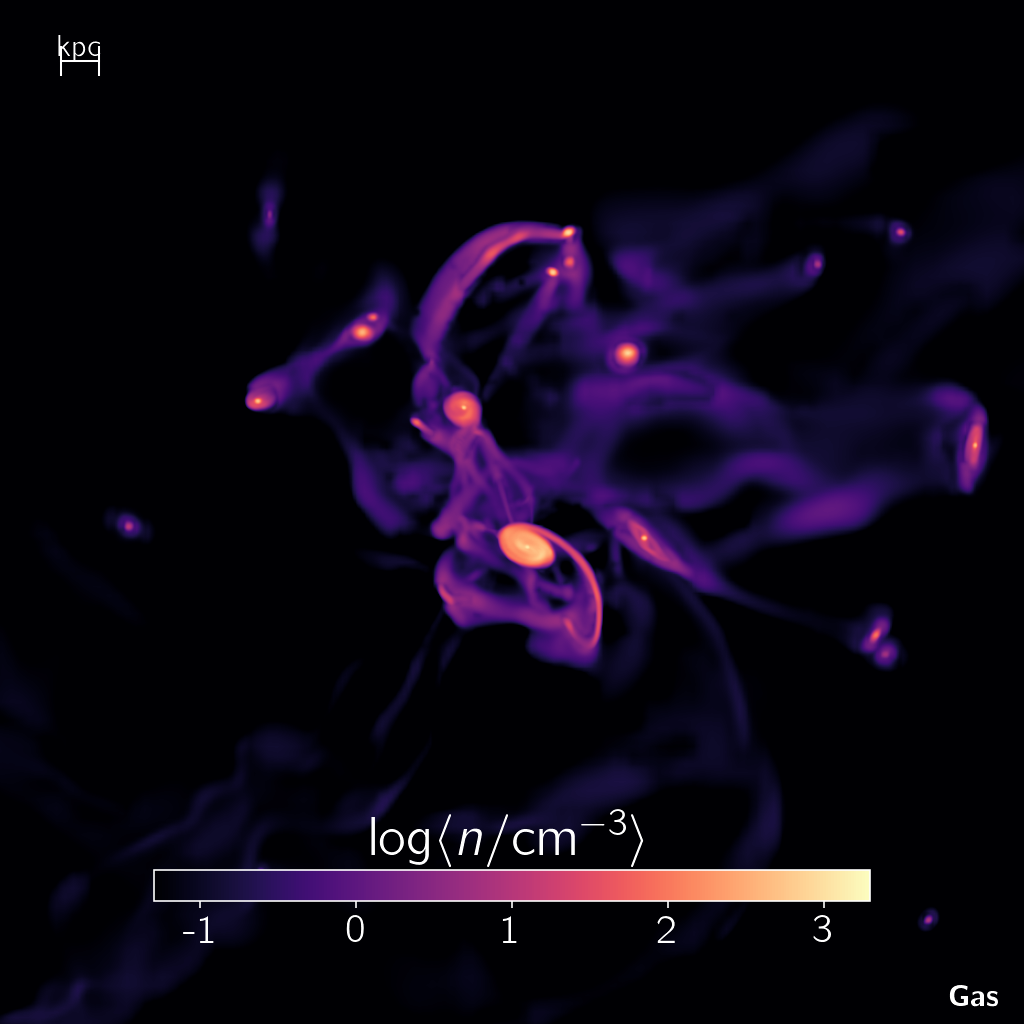}
\includegraphics[width=0.32\textwidth]{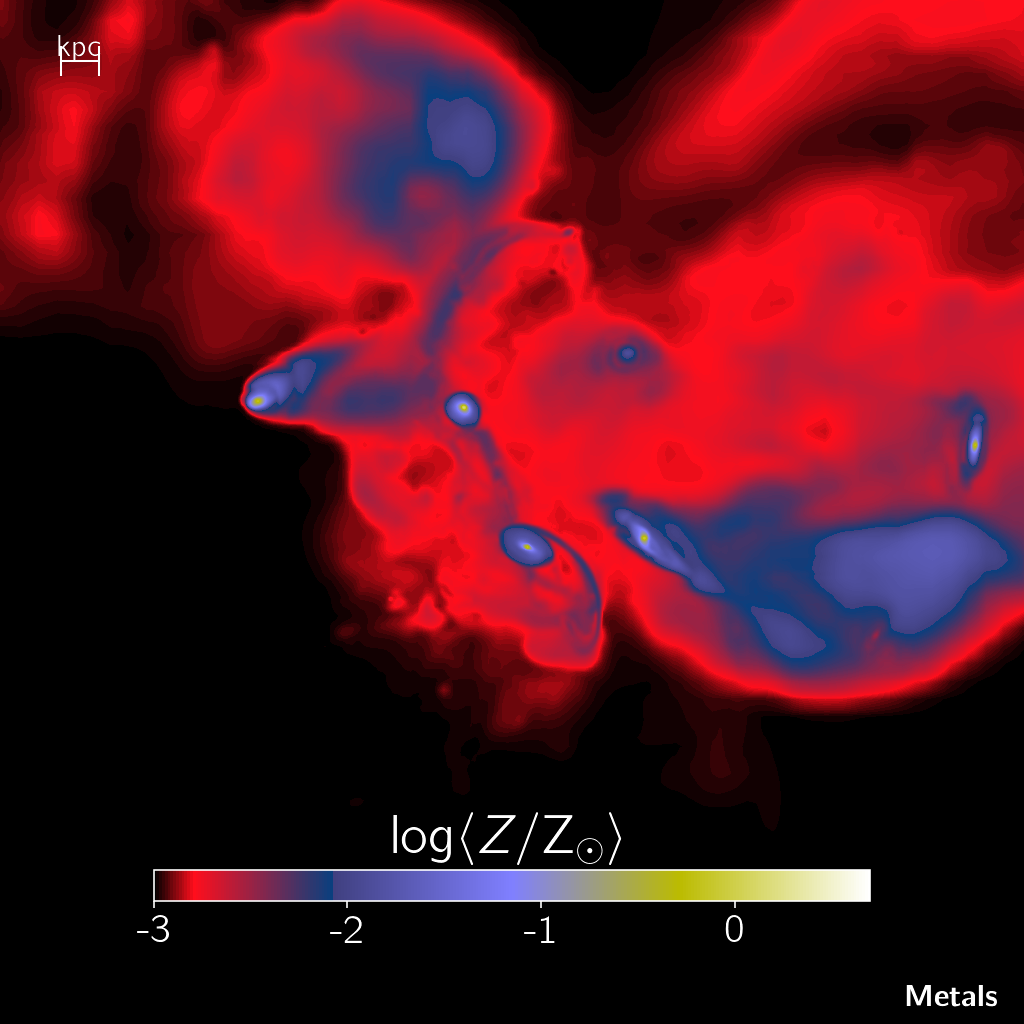}
\caption{A $z=6.3$ merging system found in the \texttt{SERRA} simulation suite  \citep{Pallottini22} selected from the \citet{kohandel:2023} sample. The central galaxy (``Adenia'') has $\log (M_\star / M_\odot) = 10.1$ and SFR$ \simeq 42 M_\odot \rm{yr}^{-1}$;
its kinematics at $z \sim 7.7$ has been analyzed in \citet{rizzo:2022}.
The physical scale of the snapshots is reported in the top-left corner of each panel (i.e. 1 kpc).
The stellar surface density ($\Sigma_\star$), average gas density ($n$) and metallicity ($z$) distributions are shown in the left, middle and right panels, respectively.
\label{fig:adenia}
}
\end{figure*}

Interestingly, even if the northern arc like structure  (C, orange histogram) has a primary photo-$z$ at $z < 1$, it shows a likelihood compatible with the redshift of the main doublet. 
The hypothesis that source C is linked to A and B is corroborated by the evidence
of a diffuse emission in the highly contrasted NIRCam image (see right panel of Fig. \ref{fig:klama-rgb}, F356W map), acting as a bridge between the main  pair (A+B) and the morphologically disturbed C component. Even if we cannot exclude
that this emission could be background noise, we consider the possibility that it is the result of a tidal interaction within the galaxy group. 

To account for the global information available from the spatially resolved methodology,
we illustrate in the bottom panel of Fig. \ref{fig:pdf} the results of a luminosity weighted 
scheme averaging the PDF(z) of all the pixels belonging to the parent object. 
It can be seen that KLAMA (A) still has the highest probability of lying at $z \sim 6$,
maintaining the degeneracy of being an \gls{LBG} at higher$z$ ($z > 10$). 
Component B maintains an almost monotonic solution at $z \sim 6$, while
component C broadens its primary peak at $z < 3$, but retains a possible (minor) secondary solution compatible with $z\sim6$.

This analysis illustrates the intrinsic level of degeneracy in the photo-$z$ derivation of optically dark sources selected through \gls{JWST} photometry, and the urgent need of spectroscopic confirmation.

In the following, we concentrate our analysis and discussion assuming that all three components of the system are at the same redshift, i.e. $z \sim 6$. This choice is mainly driven by the PDF(z) of the brightest pixels for A and B and by the presence of tidal features in the system.
To strengthen the latter hypothesis, we study in more detail the significance of the diffuse emission connecting components B and C,  already highlighted in the F356W map (Fig. \ref{fig:klama-rgb}, right panel). 
As a similar elongated light distribution appears also in other NIRCam bands, we present in Fig. \ref{fig:bridge} the S/N spatial distribution obtained by coadding the F277W, F356W and F444W images (all convolved to the same PSF of the F444W band). The color scale shows the value of S/N in each pixel.
It can be seen that the bridge connecting B and C includes many pixels with  significant flux signal (S/N$>3$). Averaging the S/N of all the pixels included in this area (see green solid contour in Fig. \ref{fig:bridge}), we obtain a of S/N$\sim2.85$. This value is not formally significant, but it probes that the diffuse emission is unlikely associated to pure noise and it might be unravelling a low surface brightness feature produced by a tidal interaction among some components of the group.

Finally, we have investigated the possibility that the KLAMA group is sitting on a larger scale structure. To this aim, we have performed a statistical analysis of the photometric catalog of NIRCam detected sources in the SMACS0723 field. As reference, we adopt the version 7 of the public catalog provided by G. Brammer at the Dawn JWST archive\footnote{https://dawn-cph.github.io/dja/imaging/v7/}. The catalog includes photometric redshifts computed with EAZY-PY\footnote{https://github.com/gbrammer/eazy-py} \citep{Brammer2008}. The derived redshift distribution does not present any significant excess in the range $5.5<z<6.5$ (i.e. roughly corresponding to the width of the primary peak of the PDF distribution around $z\sim5.7$, see Fig. \ref{fig:pdf}). 
Moreover, we verified that the closest source to the dark merger, with a compatible photo-$z$, is located at a distance of 14.5$\arcsec$ from component A. This would correspond to a physical distance of $\sim86$ Kpc. Understanding the likelihood of a gravitational connection between this source and the main system would require further dynamical analysis, that is beyond the scope of this paper.  These preliminary results, however, do not provide evidence for the existence of  a clear galaxy overdensity at the redshift of the KLAMA group.

\subsection{Physical properties of a merging system at $z = 5.7$}

We assume a value of $z = 5.7$ for the photometric redshift of the system, corresponding to the 50th percentile of the PDF($z$) of the main component A (see Fig.\,\ref{fig:pdf}).
We then run again \texttt{Bagpipes}, this time considering the same parameters as the previous run but fixing the redshift to the same values for all pixels on the map. The results for the main physical properties are shown in Fig.\,\ref{fig:physical}, where we report with different color scales the spatial distribution of stellar mass (in logarithmic scale, left panel), \gls{sfr} (in logarithmic scale, central panel) and the extinction parameterized as $A_V$. 
We can then derive our general conclusions about the properties
of a complex structure with multiple components. 

The primary object of our selection, KLAMA (A), is clearly the most luminous component in the F444W band, and, thus the most massive.
We computed the integrated values of stellar mass and \gls{sfr} by simply summing over all the pixels belonging to each source. Fig. \ref{fig:physical} shows the contours of the segmentation map (generated with \texttt{SExtractor}) used to delimit the boundaries for the integration within each object.

With $\log (M_\star / M_\odot) = 10.3 \pm 0.51$ and $\log (\mathrm{SFR} / M_\odot\,\mathrm{yr}^{-1}) = 1.91 \pm 0.49$, KLAMA is a galaxy at $z=5.7$ sitting on the main sequence of star-forming galaxies  \citep[e.g.][]{Popesso23}.
According to the best-fit model, its extremely red color is due to the dusty nature of this object, reaching values of $A_V>3$ towards the central core. To obtain an integrated value for $A_V$ we compute the average over all the pixels (see Table \ref{tab:phys}). 
The bluer components, B and C, both have negligible ($A_V<1$) dust attenuation, lower stellar masses, with $\log (M_\star / M_\odot) = 8.85 \pm 0.10$ and $\log (M_\star / M_\odot) = 8.79 \pm 0.10$, respectively. Relying on the \gls{sfr} derived from the \gls{sed} fitting approach, these values locate also the B and C components on the main sequence of star-forming galaxies at $z\sim6$ \citep{Popesso23}.

A summary of the physical properties of the three components of the system is reported in Table \ref{tab:phys}.

\begin{table}
    \caption{Integrated physical properties of the three galaxies of the system, assuming a redshift of  at $z=5.7$ (see Fig. \ref{fig:physical}).}
    \centering
    \begin{tabular}{lccc}
    \toprule
    ID & $\log M_{\star}$ & $\log\,\mathrm{SFR}$ & <$A_V$>\\
              & $M_{\odot}$ & $M_\odot$ yr$^{-1}$ & mag \\
    \hline
    \noalign{\vskip 1pt}    
    A & $10.30 \pm 0.51$  & $1.91 \pm 0.49$  & $1.85 \pm 0.72$ \\
    B & $\phantom{0}8.83 \pm 0.10$ & $0.84 \pm 0.15$  & $0.57 \pm 0.16$ \\
    C & $\phantom{0}8.93 \pm 0.23$ & $1.03 \pm 0.24$  & $0.47 \pm 0.17$ \\
   \hline
    \end{tabular}
    \label{tab:phys}
\end{table}

\section{Discussion}\label{discussion}

In this paper, we have reported the identification of a candidate multiple merging systems at $z \sim 6$, involving a massive optically-dark object with $\log (M_\star / M_\odot) \sim 10.3$, and a smaller bluer source with $\log (M_\star / M_\odot) \sim 9$. It has been suggested that these obscured objects could represent the progenitors of the largest present-day galaxies in massive groups and clusters \citep{wang19}.
It is thus crucial to monitor how their stellar mass is accumulated through the merging of galaxies in high-density environments at high redshifts, in comparison to the accretion of cold gas flows from the cosmic web \citep[e.g.][]{Keres2005,Dekel2009}.

Moreover, galaxy mergers are believed to play a significant role in stellar mass growth throughout cosmic history. They occur frequently across the timeline of the Universe and can be observed using various methods and at different evolutionary stages.
The merging of galaxies is a natural prediction of the hierarchical structure formation model
\citep{White1991,Klypin2011} for which dark matter halos grow their mass and affect the build-up of galaxies in the Universe \citep[e.g.][]{Khochfar05}. Not only can mergers increase the stellar mass of galaxies \citep[up to a factor of 2 for major mergers; e.g.][]{Lopez12,Kaviraj2014}, but they can also trigger starbursts and active galactic nuclei.
Recently, new constraints are revealing the presence of a significant merging fraction (i.e. $>30$--$40\%$) at $4 < z < 6$ \citep{Conselice09,Romano21}.

\begin{figure*}
    \centering
   \includegraphics[width=18cm]{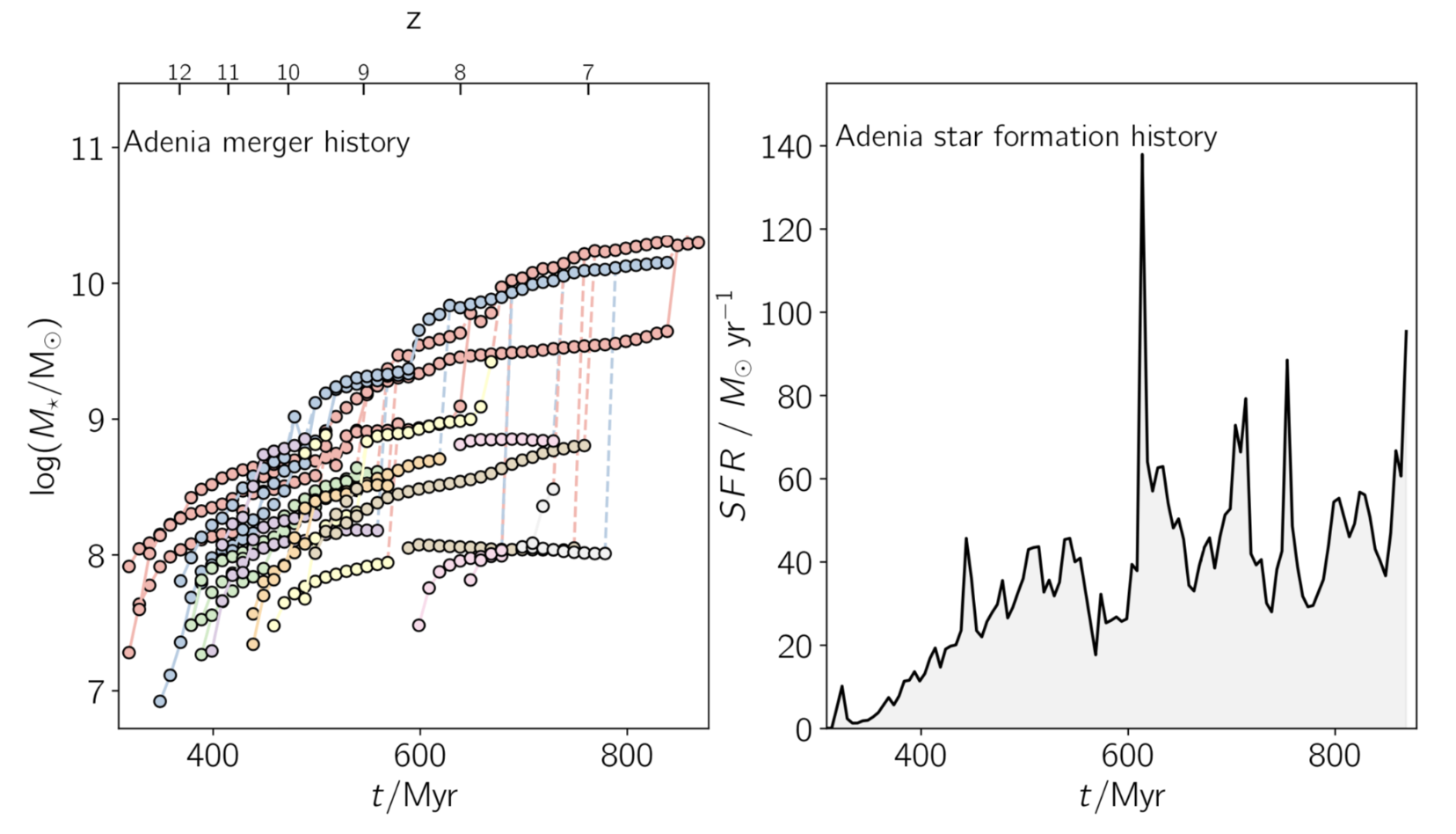}
  \caption{{\it Left panel:} Merger history of Adenia, a simulated \texttt{SERRA} galaxy similar to KLAMA. Individual lines show the galaxies' stellar mass ($M_*$) eventually merging with Adenia as a function of cosmic time ($t$, redshift $z$ is indicated in the upper axis). Each galaxy track is indicated with a different color; solid lines indicate when the different galaxies evolve as dynamical distinct systems according to the \texttt{ROCKSTAR} phase-space clump finder \citep{Behroozi13}, dashed lines highlight the merging events; see \citet{Pallottini22} for details of the merger history computation, in particular Sec. 2.4.1 therein. {\it Right panel:} Star formation history of Adenia as a function of the lookback time with respect to $z = 6.3$, when the galaxy is experiencing a merger (see Fig. \ref{fig:adenia})}.
    \label{fig:serra_mergerhistory}      
\end{figure*}

Thus, assuming that the whole observed system will eventually really merge with time, we can say that the main optically dark A component  will approximately increase its stellar mass by a $\sim 10\%$ (if B and C will both merge with A) and that on the order of other ten similar merging events are required to double its mass. To understand the origin and the frequency of merging dusty groups at the epoch of KLAMA, we have searched for similar systems within the \texttt{SERRA} simulation suite \citep{Pallottini22}.

In \texttt{SERRA}, dark matter, gas, stars are evolved with a customized version of \texttt{RAMSES} \citep{teyssier:2002}. \texttt{KROME} \citep{grassi:2014} is adopted in order to follow the non-equilibrium chemistry of up to molecular hydrogen ($H_2$) formation \citep{bovino:2016,pallottini:2017_b}. Radiation is tracked on-the-fly using \texttt{RAMSES-RT} \citep{rosdahl:2013}, that is coupled to the chemical evolution of the gas \citep{pallottini:2019,decataldo:2019}.
Adopting a \citet{schmidt:1959}-\citet{kennicutt:1998} relation, H$_2$ is converted into stars, which act as a source for mechanical energy, photons, and reprocessed elements according to the metallicity and age of the stellar population \citep{bertelli:1994}. Feedback considers thermal and kinetic energy deposition from supernovae, winds from massive stars, and an approximate treatment of radiation pressure \citep{pallottini:2017}.

Each of the simulations in the suite starts at $z = 100$ from cosmological initial conditions and zooms-in on a target DM halo ($M_h \sim 10^{12} M_\odot$ at $z = 6$) and its environment -- approximately $(2\,{\rm Mpc}/{\rm h})^{3}$ -- reaching a gas mass and spatial resolution of $\simeq 1.2 \times 10^4 M_\odot$ and $\simeq 25\,{\rm pc}$ at $z = 6.3$ in the densest regions, i.e. typical mass and size of Galactic molecular clouds.

Exploiting the \citet{kohandel:2023} sample of $\sim 3000$ synthetic observations of \texttt{SERRA} galaxies between $4 \lesssim z \lesssim 9$ \citep[see][for the generation of the data cubes]{kohandel:2020}, we identify a merging system with physical properties similar to those of the observed one where KLAMA resides.
Such a simulated system is presented in Figure \ref{fig:adenia}, where we report maps of the stellar surface density, gas density, and metallicity (in the left, middle and right panels, respectively). The central object is ``Adenia'', a $z = 6.3$ galaxy with a stellar mass of $\log (M_\star / M_\odot) = 10.1$, $\log (\mathrm{SFR} / M_\odot\,\mathrm{yr}^{-1}) = 1.62$, and an average $A_V = 0.7$ mag that reaches values as high as $A_V \sim 5$--$6$ mag in the central 0.7 kpc (corresponding to the physical scale of the innermost beam of KLAMA, see Fig.\,\ref{fig:physical}).
All these values, and the presence of a companion source at a physical distance of $\sim$4 kpc, similar to that of components A and B in Fig \ref{fig:klama-rgb} ($\sim$3 kpc), showcase that sources as KLAMA are predicted to exist in state-of-the-art cosmological simulations.

The gas density distribution (middle panel of Fig. \ref{fig:adenia}) highlights the presence of gas streams as the fossil record of the gravitational interactions that involved Adenia in a spherical volume with a $\sim 10$ kpc diameter.
This view allows us to better catch the similarity of the simulation with the observation. 
Indeed, the potential tidal features bridging the "tail" and the "head" of the Whale system (see right panel of Fig. \ref{fig:klama-rgb}) are explained by the complex dynamical interactions among the sources of the group in the simulation.

We have studied Adenia's past merger history to infer the stellar mass build-up of KLAMA. This is presented in Figure \ref{fig:serra_mergerhistory} (left panel), where the merger tree of the central simulated \texttt{SERRA} galaxy, now at $z\simeq6.3$, is shown. Multiple mergers have contributed to the stellar mass growth of Adenia, the latest being, at the age of $\sim 848$ Myr, between two massive galaxies of mass ratio $\sim 10$. This is intriguingly similar to the stellar mass ratio of the observed component A with respect to the less massive B+C objects. On the right panel of \ref{fig:serra_mergerhistory}, we also report the star formation history of Adenia, showing different starburst events likely associated with a few dramatic merger events.
Furthermore, the physical properties of the interstellar medium of Adenia (characterized by high levels of obscuration) allow for the prediction of its observability at submillimeter wavelengths. In particular, the [CII] total luminosity of the central galaxy in the simulation is $L_{\rm CII} \sim 1.7 \times 10^8 L_{\odot}$ with FWHM $\sim 530\,{\rm km}\,{\rm s}^{-1}$.
For a firm detection with the Atacama Large Millimeter Array (ALMA), it would take $\sim 10-25$ hours of observation,  depending on the inclination (face-on/edge-on) of the system, without accounting for the need of spectral scans, making it challenging but feasible.

\begin{figure}
    \centering
   \includegraphics[width=9cm]{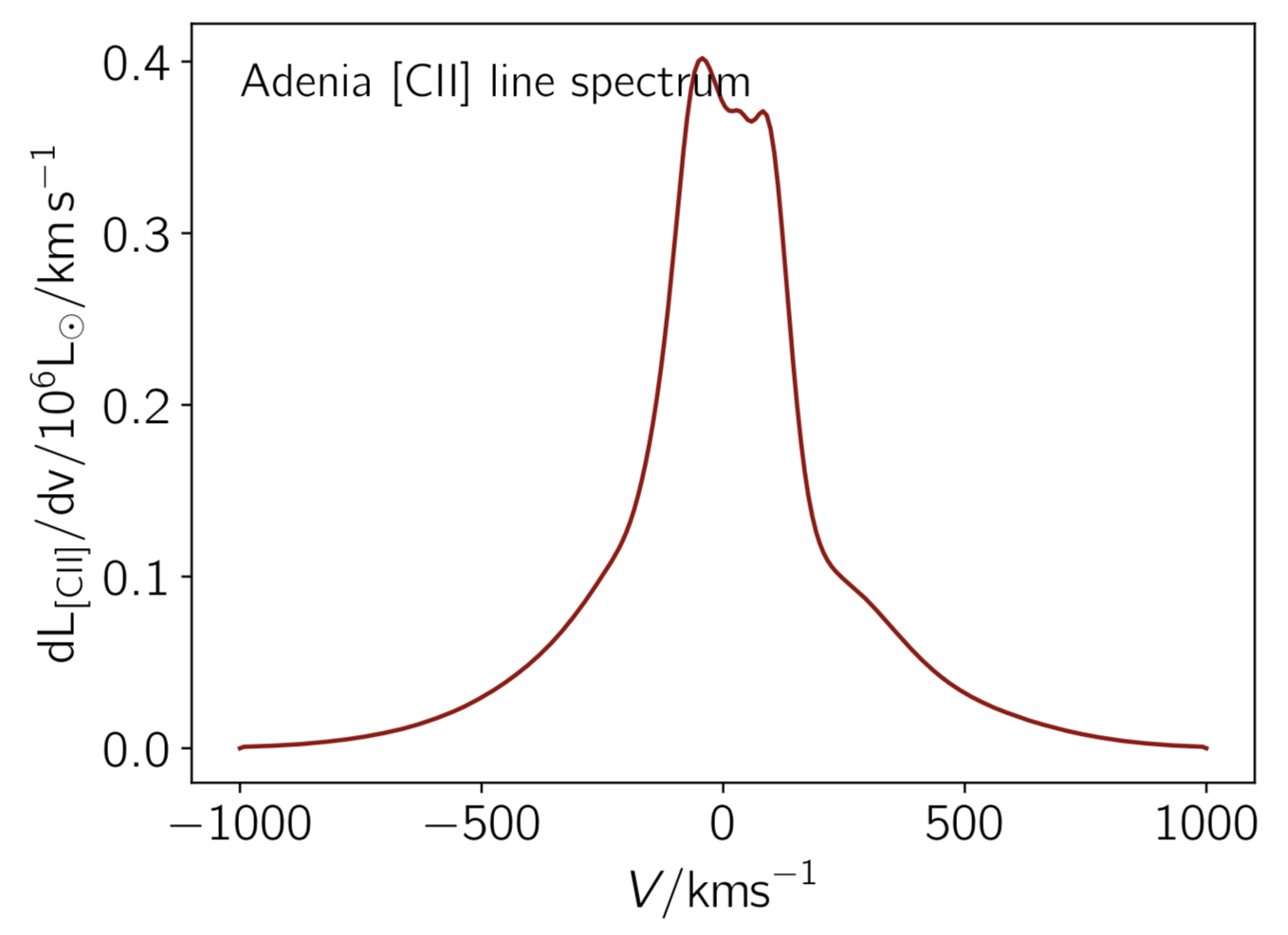}
 \caption{Synthetic [CII] spectra from the galaxy Adenia. The spectrum shows an integrated luminosity of $L_{[CII]}\simeq 1.7 \times 10^8 {\rm L}_\odot$ and full width half maximum of $\simeq 530 \,\rm km\,s^{-1}$. The line emission has been computed from the simulated galaxy by using cell-by \texttt{CLOUDY} models \citep{ferland:2017} accounting for the local interstellar radiation field, the gas properties (density, metallicity, proper motion, etc), and the turbulent structure of molecular clouds \citep{vallini:2018, pallottini:2019, kohandel:2019}. 
 \label{fig:CIIline}
 }        
\end{figure}

The overall physical properties of KLAMA also suggest that this system could be similar to the well-known submillimeter galaxy (SMG) HFD850.1 \citep{Hughes}, a heavily dust-obscured merging galaxy at $z = 5.18$ \citep{Neri}. The stellar component of the source has been recently resolved by NIRCam \citep{Sun2024}, revealing a stellar mass of $\log (M_\star / M_\odot) = 10.8 \pm 0.1$. HFD850.1 is among the first discoveries of optical/near-IR-dark SMGs, and it is sitting in a well-characterized galaxy overdense region, suggesting that up to $\sim 50\%$ of the cosmic \gls{SFRD} at $z \sim 5.0$--$5.5$ could be contributed by sources in protocluster structures \citep[][and references therein]{Sun2024}. It is thus natural to complement the submillimeter search for elusive high-$z$ obscured groups and protoclusters with deep near-IR surveys, currently achievable with the \gls{JWST}.  

The conclusions of this work are based on just one object, and further statistical investigation is needed to obtain a definitive answer about the role of merging in the stellar mass assembly of dusty system at $z\sim5-6$.
However, our results (based on the comparison of simulations with observations) put strong evidence on the fact that mergers are an essential driver of the galaxy's build up during the obscured star formation phase, at least up to the end of the Epoch of Reionization.

\section{Summary}
In this paper, we have presented and discussed a single case of a massive, optically dark source (selected through its red F277W-F444W color) likely undergoing a merging process. 
The detection of these interactions within a dusty galaxy group, occurring right at the conclusion of the Epoch of Reionization, directs focus towards the significance of mergers in the assembly of stellar mass and their impact on the cosmic \gls{sfr} at high redshifts. 
We summarize below the main findings:
\begin{itemize}
\item We looked for companions in a sample of ODGs in the SMACS0723 field. The brightest F444W source (with an integrated magnitude on the order of [F444W]$\sim$25mag) shows a close bluer companion at a distance of $<0.5"$, and a secondary elongated object (potentially showing multiple components) at $\sim1.5"$. The latter is connected to the southern pairs by some tidal features (making the morphology of the group similar to that of a whale).
\item Exploiting the multiwavelength NIRCam observations in this sky region, we convolved each image to the lowest angular resolution (i.e. that of the F444W band). We then performed a spatially resolved SED fitting on each PSF-matched pixel to derive the physical properties of this galaxy system.
\item We found that all the components of the group are consistent with a photometric redshift of $z\sim5.7$. The darkest object turns out to be a massive and dusty galaxy,
with $\log (M_\star / M_\odot) = 10.3 \pm 0.51$, $\log (\mathrm{SFR} / M_\odot\,\mathrm{yr}^{-1}) = 1.91 \pm 0.49$
and $A_V\sim3$ in the central reddest portion of the object.
The group's other sources are bluer and show negligible dust attenuation. The mass ratio between the more massive source and the other surrounding companions is on the order of $1/10$.
The physical spatial extension of the whole system is $\sim10$kpc at $z=5.7$.
\item We finally investigated the lightcones in the \texttt{SERRA} cosmological simulation, and we identified a source showing the same physical and environmental properties of the Whale group presented in this paper (including the presence of companions with the same mass ratio of the observed system within a $\sim10$kpc physical distance). The simulation thus allowed us to infer the stellar mass assembly history of a massive, dusty, optically dark source, showing that multiple merger events are required to build up such an evolved system in times as early as the end of the EoR. Such objects could be directly studied through their [CII] emission line.

\item The conclusions drawn from this study rely on a single object, indicating the need for further statistical investigation to definitively determine the role of merging in the stellar mass assembly of dusty systems at $z\sim5-6$. Nevertheless, our findings, supported by comparisons between simulations and observations, strongly suggest that mergers play a crucial role in galaxy formation during the obscured star formation phase, at least until the end of the Epoch of Reionization.

\end{itemize}

We caution that, in order to provide a quantitative measurement of the role of the local environment in the most obscured and massive system from Cosmic Dawn to Cosmic Noon, it is mandatory to perform a systematic analysis in statistical samples (in both observations and simulations). 
In a future paper (Girardi et al. in prep.), we will perform an in depth search for physical companions of optically dark sources across all available extragalactic \gls{JWST} fields.

We also remind the reader that a spectroscopic confirmation is required to unravel the nature and the redshift of the KLAMA system. While submillimeter observations with ALMA  have already identified mergers in obscured systems at $z\sim5$--$6$ \citep[e.g.][]{Romano21}, \gls{JWST} represents the ideal instrument to resolve the KLAMA puzzle (in particular the NIRSpec Integral Field Unit that perfectly matches the field of view of our target, see left panel of Fig.\,\ref{fig:klama-rgb}).

\begin{acknowledgements}
G.R., B.V., G.G. and A.G. are supported by the European Union – NextGenerationEU RFF M4C2 1.1 PRIN 2022 project 2022ZSL4BL INSIGHT.
G.R., L.B. and A.E. also acknowledge the support from MIUR grant PRIN 2017 20173ML3WW-0013.
A.B. and G.R. acknowledge support from INAF under the Large Grant 2022 funding scheme (project ``MeerKAT and LOFAR Team up: a Unique Radio Window on Galaxy/AGN co-Evolution”).
B.V. acknowledges support from the INAF Mini Grant 2022 “Tracing filaments through cosmic time”  (PI Vulcani). 
We also  acknowledge support from the INAF Large Grant 2022 “Extragalactic Surveys with JWST” (PI Pentericci).
M.K., A.P., and A.F. acknowledge the CINECA award under the ISCRA initiative for the availability of high-performance computing resources and support from the Class B project SERRA HP10BPUZ8F (PI Pallottini).
M.K., A.P., and A.F. gratefully acknowledge the computational resources of the Center for High-Performance Computing (CHPC) at SNS.
A.F. acknowledges support from the ERC Advanced Grant INTERSTELLAR H2020/740120 (PI Ferrara). Any dissemination of results must indicate that it reflects only the author's view and that the Commission is not responsible for any use that may be made of the information it contains. Partial support (A.F.) from the Carl Friedrich von Siemens-Forschungspreis der Alexander von Humboldt-Stiftung Research Award is kindly acknowledged.
\end{acknowledgements}

\bibliographystyle{aa}
\bibliography{bib.bib} 

\end{document}